\documentclass[a4paper,11pt]{article}
\usepackage{pos}
\usepackage{mathtools}
\usepackage{bm}
\usepackage{siunitx}
\usepackage{slashed}
\usepackage{physics2}
\usephysicsmodule{braket}
\usepackage{tabularray}
\usepackage{float}
\DeclareMathOperator{\Tr}{Tr}

\begin{document}
\title{Hadronic Molecules and $\chi_{cJ}(2P)$ Coupled States}
\author*[a]{Kotaro Miyake}
\author[a,b]{Yasuhiro Yamaguchi}
\affiliation[a]{Department of Physics, Nagoya University, Nagoya 464-8602, Japan}
\affiliation[b]{Kobayashi-Maskawa Institute for the Origin of Particles and the Universe, Nagoya University, Nagoya, 464-8602, Japan}
\emailAdd{miyake@hken.phys.nagoya-u.ac.jp}
\emailAdd{yamaguchi@hken.phys.nagoya-u.ac.jp}
\abstract{We investigate hidden-charm exotic mesons based on a coupled-channel framework, where the physical states are described as superpositions of a compact $c\bar c$ core (identified with the $\chi_{cJ}(2P)$ states) and hadronic components such as $D^{(*)}\bar D^{(*)}$. We incorporate meson exchange potentials to describe hadron-hadron interactions and determine model parameters to reproduce the observed masses of $X(3872)$ and $Z(3930)$. The model also predicts a $0^{++}$ bound state consistent with the $X(3860)$. The internal structure is found to be predominantly molecular for the $X(3872)$, while the $X(3860)$ and $Z(3930)$ have larger $c\bar{c}$ components. The coupling between $c\bar{c}$ cores and hadronic components plays a crucial role in generating these states. The resulting wave functions provide insight into the internal structure and decay properties of these exotic states.}
\FullConference{The 21st International Conference on Hadron Spectroscopy and Structure (HADRON2025)\\
	27 - 31 March, 2025\\
	Osaka University, Japan\\}
\maketitle

\section{Introduction}\label{sec:intro}
The discovery of exotic hadrons that deviate from the conventional quark model has stimulated intense interest in hadron physics. In particular, many candidates containing charm quarks—such as the $X$, $Y$, $Z$, $T_{cc}$, and $P_c$ states—have been reported in recent experiments~\cite{Brambilla:2019esw,Yamaguchi:2019vea,Chen:2022asf,ParticleDataGroup:2024cfk}. These states often appear near relevant hadronic thresholds, suggesting molecular interpretations as loosely bound states of conventional hadrons. However, the mechanisms driving such binding and the internal structures of these states remain elusive.

Among them, the $X(3872)$\cite{Belle:2003nnu} is a well-established state located very close to the $D^0\bar{D}^{*0}$ threshold. Its quantum numbers, decay patterns, and prompt production behavior suggest that it is not a simple hadronic molecule nor a pure charmonium, but rather a superposition of hadronic and compact $c\bar{c}$ components~\cite{Esposito:2015fsa,Yamaguchi:2019vea}. In particular, a possible $c\bar{c}$ component is identified with the yet-unobserved $\chi_{c1}(2P)$ state predicted by constituent quark models~\cite{Godfrey:1985xj}.

Other $X$ states such as $X(3860)$ ($0^{++}$) and $Z(3930)$ ($2^{++}$) are regarded as spin partners of the $X(3872)$ and also appear near $D^{(*)}\bar{D}^{(*)}$ thresholds. The proximity to the predicted $\chi_{cJ}(2P)$ charmonium masses motivates a unified treatment of these states within a coupled-channel framework.

In this work, we investigate hidden-charm exotic mesons using a coupled-channel framework, where the physical states are described as superpositions of a compact $c\bar{c}$ core (identified with the $\chi_{cJ}(2P)$ states) and $D^{(*)}\bar{D}^{(*)}$ hadronic components~\cite{Miyake:2025ktz}. We incorporate meson exchange potentials to describe hadron-hadron interactions and determine model parameters to reproduce the observed masses of $X(3872)$ and $Z(3930)$. Our calculations also produce a $0^{++}$ bound state whose mass is consistent with that of the $X(3860)$.  The resulting wave functions provide insight into the internal structure of these exotic states.

\section{Formalism}
\subsection{Schrödinger Equation}
We investigate the $\chi_{cJ}(2P)$ states that can be observed in experiments, modeling them as superpositions of bare $\chi_{cJ}(2P)$ states and hadronic molecule components. For the latter, we include the $D^{(*)}\bar{D}^{(*)}$ and $D_s \bar{D}_s$ channels listed in Table~\ref{tab:channels}, excluding $D\bar{D}$ and $D\bar{D}^*$ for $0^{++}$ and $2^{++}$ due to their limited contributions. $G$-wave components are also neglected for simplicity.
\begin{table}
	\centering
	\caption{\label{tab:channels}Channels considered in the hadronic molecule part {for given $J^{PC}$}. $[D\bar{D}^{*}]$ denotes $\frac{1}{\sqrt{2}}(D\bar{D}^{*}-D^{*}\bar{D})$.}
	\begin{tblr}{X[1,c]|X[10,c]}
		$J^{PC}$ & Channel                                                                                                                                                                                                 \\
		\hline
		$0^{++}$ & $D_s^+D_s^-(^1S_0)$,\quad$D^{*0}\bar{D}^{*0}(^1S_0)$,\quad$D^{*+}D^{*-}(^1S_0)$,\quad$D^{*0}\bar{D}^{*0}(^5D_0)$,\quad$D^{*+}D^{*-}(^5D_0)$                                                             \\
		\hline
		$1^{++}$ & $[D^0\bar{D}^{*0}](^3S_1)$,\quad$[D^+D^{*-}](^3S_1)$,\quad$[D^0\bar{D}^{*0}](^3D_1)$,\quad$[D^+D^{*-}](^3D_1)$,\quad$D^{*0}\bar{D}^{*0}(^5D_1)$,\quad$D^{*+}D^{*-}(^5D_1)$                              \\
		\hline
		$2^{++}$ & $D_s^+D_s^-(^1D_2)$,\quad$D^{*0}\bar{D}^{*0}(^5S_2)$,\quad$D^{*+}D^{*-}(^5S_2)$,\quad$D^{*0}\bar{D}^{*0}(^1D_2)$,\quad$D^{*+}D^{*-}(^1D_2)$,\quad$D^{*0}\bar{D}^{*0}(^5D_2)$,\quad$D^{*+}D^{*-}(^5D_2)$
	\end{tblr}
\end{table}

The coupled-channel Schrödinger equation is given by
\begin{align}
	\mathcal{H}\Psi & = E\Psi, \\
	\mathcal{H}     & =
	\begin{pmatrix}
		H_0 + V_\mathrm{OBE} & \mathcal{U}^\dagger                  \\
		\mathcal{U}          & m_{\chi_{cJ}} - m_\mathrm{threshold}
	\end{pmatrix},
\end{align}
where $H_0$ is the kinetic term, and $V_\mathrm{OBE}$ is the one-boson-exchange (OBE) potential. The wave function $\Psi$ includes both hadronic molecule states and the bare state. The transition potential $\mathcal{U}$ accounts for mixing between them.

\subsection{Heavy Meson Chiral Lagrangian}
$V_\mathrm{OBE}$ is derived from heavy meson chiral Lagrangians, respecting heavy quark spin symmetry, chiral symmetry, and Lorentz invariance.
The linear combination $H^{(Q)}$ of the pseudoscalar field $P$ and vector field $P^*$ and its charge conjugates $H^{(\bar{Q})}$ are defined as
\begin{gather}
	H^{(Q)}_a \equiv \frac{1+\slashed{v}}{2}(P^{*\mu}_a\gamma_\mu-P_a\gamma^5),\quad\bar{H}^{(Q)}_a \equiv \gamma^0(H^{(Q)}_a)^\dagger\gamma^0 = (P^{*\mu\dagger}_a\gamma_\mu+P^\dagger_a\gamma^5)\frac{1+\slashed{v}}{2}, \\
	H^{(\bar{Q})}_a  \equiv C(\mathcal{C}H^{(Q)}_a\mathcal{C}^{-1})^{\top}C^{-1}  = (\bar{P}^{*\mu}_a\gamma_\mu-\bar{P}_a\gamma^5)\frac{1-\slashed{v}}{2},\quad\bar{H}^{(\bar{Q})}_a = \frac{1-\slashed{v}}{2}(\bar{P}^{*\mu\dagger}_a\gamma_\mu+\bar{P}^\dagger_a\gamma^5),
\end{gather}
where $v^\mu$ is the four-velocity of the heavy meson.
The interaction Lagrangians are constructed to describe both pseudoscalar and vector meson exchanges. The pseudoscalar meson exchange is described by
\begin{equation}
	\mathcal{L}_{MHH} = ig\Tr{\left[H^{(Q)}_b\gamma^\mu\gamma^5A_{ba\mu}\bar{H}^{(Q)}_a\right]} + ig\Tr{\left[\bar{H}^{(\bar{Q})}_a\gamma^\mu\gamma^5A_{ab\mu}H^{(\bar{Q})}_b\right]},
\end{equation}
while the vector meson exchange is described by
\begin{multline}
	\mathcal{L}_{VHH} = i\beta\Tr{\left[H^{(Q)}_b v_\mu(V^\mu_{ba}-\rho^\mu_{ba})\bar{H}^{(Q)}_a\right]}
	+i\lambda\Tr{\left[H^{(Q)}_b\sigma_{\mu\nu}F^{\mu\nu}_{ba}\bar{H}^{(Q)}_a\right]}                    \\
	-i\beta\Tr{\left[\bar{H}^{(\bar{Q})}_a v_\mu(V^\mu_{ab}-\rho^\mu_{ab})H^{(\bar{Q})}_b\right]}
	+i\lambda\Tr{\left[\bar{H}^{(\bar{Q})}_a\sigma_{\mu\nu}F^{\prime\mu\nu}_{ab}H^{(\bar{Q})}_b\right]},
\end{multline}
where $A_{\mu}$ ($V_\mu$) is the axial (vector) current of light pseudoscalar mesons, and $\rho^\mu$ is the light vector meson field. The parameters $g$, $\beta$, and $\lambda$ are coupling constants determined from experimental data or lattice QCD calculations.

\subsection{OBE Potential}
We include $\pi$, $\eta$, $K$, $\rho$, $\omega$, $K^*$, and $\phi$ exchanges. In calculating the potential, we apply the static approximation and neglect short-range delta-function terms. The potential includes central and tensor parts, regularized by a monopole-type form factor:
$F(\bm{q},m) = {\left(\frac{\Lambda^2 - m^2}{\Lambda^2 + \bm{q}^2}\right)^2}$
with $\Lambda = \alpha \times \qty{220}{MeV} + m$. We vary $\alpha$ to assess cutoff dependence.

\subsection{Transition Potential}
The transition potential $\mathcal{U}$ connects the bare $\chi_{cJ}(2P)$ state with $S$-wave hadronic molecules. We adopt the model of Refs.~\cite{Takizawa:2012hy,Miyake:2025ktz}, where the spatial form is given by a Yukawa-type function with parameters $g_{c\bar{c}}$ and $\Lambda_q$. Spin factors $f_\mathrm{spin}$ are determined by Clebsch-Gordan coefficients and listed in Table\ref{tab:spin-factor}.
\begin{table}
	\caption{\label{tab:spin-factor}Spin factor $f_\mathrm{spin}$ for each channel.}
	\centering
	\begin{tabular}{ccccc}
		\hline\hline
		                  & $D_s^+D_s^-(^1S_0)$ & $D^{*}\bar{D}^{*}(^1S_0)$ & $D\bar{D}^{*}(^3S_1)$ & $D^{*}\bar{D}^{*}(^5S_2)$ \\
		\hline
		$f_\mathrm{spin}$ & $\frac{1}{2}$       & $\frac{\sqrt{3}}{2}$      & $1$                   & $1$                       \\
		\hline\hline
	\end{tabular}
\end{table}

\section{Numerical Results}\label{sec:result}

We investigate bound states of the physical $\chi_{cJ}(2P)$ by solving the coupled-channel Schrödinger equation. The OBE coupling constants are fixed using experimental and lattice QCD inputs, as summarized in Table~\ref{tab:free-parameters}. The bare masses $m_{\chi_{cJ}}$ are taken from the Godfrey-Isgur (GI) model~\cite{Godfrey:1985xj,Barnes:2005pb} (Table~\ref{tab:chi_mass}).
\begin{table}
	\caption{\label{tab:free-parameters} {Parameters $g_{c\bar{c}}$ and $\Lambda_q$ determined to reproduce the masses of $X(3872)$ and $Z(3930)$ for given $\alpha=0.7$, $1.0$ and $1.3$, respectively.}}
	\centering
	\begin{tabular}{cccc}
		\hline\hline
		$\alpha$                & $0.7$    & $1.0$    & $1.3$    \\
		$g_{c\bar{c}}$          & $0.0448$ & $0.0427$ & $0.0409$ \\
		$\Lambda_q$(\unit{MeV}) & $2260$   & $3089$   & $4647$   \\
		\hline\hline
	\end{tabular}
\end{table}
\begin{table}
	\centering
	\caption{\label{tab:chi_mass}The masses of the bare $\chi_{cJ}(2P)$ state}
	\begin{tabular}{ccc}
		\hline\hline
		$\chi_{c0}$     & $\chi_{c1}$     & $\chi_{c2}$     \\
		\hline
		\qty{3916}{MeV} & \qty{3953}{MeV} & \qty{3979}{MeV} \\
		\hline\hline
	\end{tabular}
\end{table}

\begin{figure}
	\centering
	\includegraphics[width=0.4\textwidth]{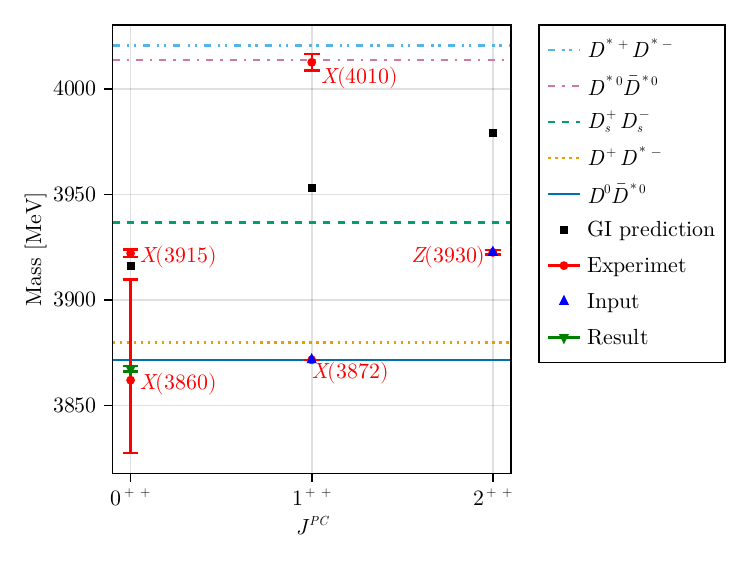}
	\caption{\label{fig:mass-of-x(3860)}Calculated mass of bound state of $0^{++}$ at each $\alpha$~\cite{Miyake:2025ktz}. {The red dots with an error are the empirical mass spectra of the tetraquarks. The blue triangles are the computed masses reproducing the masses of $X(3872)$ and $Z(3930)$. The green inverted triangle with an error is the computed mass of the $0^{++}$ bound state, where the error indicates the distribution of the obtained masses for each $\alpha$. The horizontal lines show threshold energies of $D^{(*)}_s\bar{D}^{(*)}_s$. The solid squares are mass spectra of the bare $\chi_{cJ}(2P)$, $m_{\chi_{cJ}}$, predicted by the GI model~\cite{Godfrey:1985xj,Barnes:2005pb}. }}
\end{figure}

Three free parameters—$\alpha$, $g_{c\bar{c}}$, and $\Lambda_q$—are constrained by fitting the experimental masses of $X(3872)$ ($1^{++}$) and $Z(3930)$ ($2^{++}$). We vary $\alpha = 0.7, 1.0, 1.3$ and determine $g_{c\bar{c}}$ and $\Lambda_q$ accordingly (Table~\ref{tab:free-parameters}). The mass of the $0^{++}$ state is then compared with the experimental mass of the $X(3860)$. Results are listed in Table~\ref{tab:mass-of-x(3860)} and shown in Figure~\ref{fig:mass-of-x(3860)}.
\begin{table}
	\centering
	\caption{\label{tab:mass-of-x(3860)}Calculated mass of bound state of $0^{++}$ at each $\alpha$~\cite{Miyake:2025ktz}.}
	\begin{tabular}{cccc}
		\hline\hline
		$\alpha$           & $0.7$     & $1.0$     & $1.3$     \\
		Mass\,(\unit{MeV}) & $3868.62$ & $3867.31$ & $3866.07$ \\
		\hline\hline
	\end{tabular}
\end{table}

The predicted $0^{++}$ state aligns with Belle's observation of $X(3860)$\cite{Belle:2017egg}, though not confirmed by LHCb\cite{LHCb:2020pxc}. Notably, the mass appears naturally from a model tuned only to $X(3872)$ and $Z(3930)$. Mixing ratios (Table~\ref{tab:mixing-ratio-0}) show that the $0^{++}$ state is predominantly a compact $\chi_{c0}(2P)$, with minor molecular components.
For the $1^{++}$ state (Table~\ref{tab:mixing-ratio-1}), $X(3872)$ is mostly a $D^0\bar{D}^{*0}({}^3S_1)$ molecule with small bare $\chi_{c1}(2P)$ mixing. Isospin breaking emerges due to the mass proximity to the $D^0\bar{D}^{*0}$ threshold.
The $2^{++}$ state (Table~\ref{tab:mixing-ratio-2}) shows a structure similar to $0^{++}$, dominated by the compact $\chi_{c2}(2P)$ state.

\begin{table}[H]
	\centering
	\caption{\label{tab:mixing-ratio-0}Mixing ratio of {the $0^{++}$ bound state for $\alpha=1.0$~\cite{Miyake:2025ktz}.}}
	\begin{tblr}{X[c]|X[c]X[c]X[c]X[c]X[c]X[c]}
		\scriptsize Channel      & \scriptsize $D_s^+{D}_s^{-}({}^1S_0)$ & \scriptsize $D^{*0}\bar{D}^{*0}({}^1S_0)$ & \scriptsize $D^{*+}D^{*-}({}^1S_0)$ & \scriptsize $D^{*0}\bar{D}^{*0}({}^5D_0)$ & \scriptsize $D^{*+}D^{*-}({}^5D_0)$ & \scriptsize bare $\chi_{c0}(2P)$ \\
		\hline
		\scriptsize Mixing ratio & \qty{0.66}{\percent}                  & \qty{2.53}{\percent}                      & \qty{2.45}{\percent}                & \qty{0.00}{\percent}                      & \qty{0.00}{\percent}                & \qty{94.36}{\percent}            \\
	\end{tblr}
\end{table}
\begin{table}[H]
	\centering
	\caption{\label{tab:mixing-ratio-1}Mixing ratio of {the $1^{++}$ bound state for $\alpha=1.0$~\cite{Miyake:2025ktz}.}}
	\begin{tblr}{X[c]|X[c]X[c]X[c]X[c]X[c]X[c]X[c]}
		\scriptsize Channel      & \scriptsize $D^{0}\bar{D}^{*0}({}^3S_1)$ & \scriptsize $D^+D^{*-}(^3S_1)$ & \scriptsize $D^{0}\bar{D}^{*0}(^3D_1)$ & \scriptsize $D^+D^{*-}(^3D_1)$ & \scriptsize $D^{*0}\bar{D}^{*0}(^5D_1)$ & \scriptsize $D^{*+}{D}^{*-}(^5D_1)$ & \scriptsize bare  $\chi_{c1}(2P)$ \\
		\hline
		\scriptsize Mixing ratio & \qty{82.03}{\percent}                    & \qty{4.97}{\percent}           & \qty{0.01}{\percent}                   & \qty{0.01}{\percent}           & \qty{0.00}{\percent}                    & \qty{0.00}{\percent}                & \qty{12.99}{\percent}             \\
	\end{tblr}
\end{table}
\begin{table}[H]
	\centering
	\caption{\label{tab:mixing-ratio-2}Mixing ratio of {the $2^{++}$ bound state for $\alpha=1.0$~\cite{Miyake:2025ktz}.}}
	\begin{tblr}{X[c,1.2]|X[c]X[c]X[c]X[c]X[c]X[c]X[c]X[c]}
		\scriptsize Channel      & \scriptsize $D_s^+{D}_s^{-}({}^1D_2)$ & \scriptsize $D^{*0}\bar{D}^{*0}({}^5S_2)$ & \scriptsize $D^{*+}{D}^{*-}({}^5S_2)$ & \scriptsize $D^{*0}\bar{D}^{*0}({}^1D_2)$ & \scriptsize $D^{*+}{D}^{*-}({}^1D_2)$ & \scriptsize $D^{*0}\bar{D}^{*0}({}^5D_2)$ & \scriptsize $D^{*+}{D}^{*-}({}^5D_2)$ & \scriptsize bare $\chi_{c2}(2P)$ \\
		\hline
		\scriptsize Mixing ratio & \qty{0.00}{\percent}                  & \qty{3.67}{\percent}                      & \qty{3.50}{\percent}                  & \qty{0.00}{\percent}                      & \qty{0.00}{\percent}                  & \qty{0.00}{\percent}                      & \qty{0.00}{\percent}                  & \qty{92.83}{\percent}            \\
	\end{tblr}
\end{table}

\section{Summary}
Exotic hadrons have emerged as a central topic in hadron physics, particularly the hidden-charm tetraquark candidates $X$, $Y$, and $Z$ following the discovery of the $X(3872)$ in 2003. In this study, we perform a unified analysis of the $X(3872)$ ($J^{PC}=1^{++}$) and its spin partners, $X(3860)$ ($0^{++}$) and $Z(3930)$ ($2^{++}$), based on a coupled-channel model incorporating both compact $c\bar{c}$ components and hadronic molecules such as $D^{(*)}\bar{D}^{(*)}$. The $c\bar{c}$ part is described by the $\chi_{cJ}(2P)$ states predicted in the constituent quark model.

Meson-meson interactions are constructed using effective Lagrangians with heavy quark spin and chiral symmetries, including both pseudoscalar and vector meson exchanges. Model parameters are fixed to reproduce the observed masses of $X(3872)$ and $Z(3930)$. The coupled-channel Schrödinger equation predicts a $0^{++}$ bound state consistent with the $X(3860)$.

Wave function analysis shows that the $X(3872)$ molecular components dominate, whereas the $X(3860)$ and $Z(3930)$ have larger $c\bar{c}$ fractions. The mixing between $c\bar{c}$ cores and molecular components plays a key role in forming these bound states.

Our results provide insights into the internal structure and decay properties of physical $\chi_{cJ}(2P)$ states, relevant for processes such as $\chi_{cJ}(2P) \to J/\psi V$ ($V=\gamma,\rho,\omega$) and $\chi_{cJ}(2P) \to D\bar{D}$. Extension of this model to resonant states, such as $X(3915)$ and $X(4010)$, is left for future work.

\acknowledgments
This work is supported by the RCNP Collaboration Research Network program as the project number COREnet-056.
\bibliographystyle{JHEP}
\bibliography{references}
\end{document}